\documentclass[11pt,letterpaper]{article}
\usepackage{jheppub}
\usepackage{amsmath}
\usepackage{amssymb}
\usepackage{graphicx}
\usepackage{caption}
\usepackage{subcaption}
\usepackage{dsfont}
\usepackage{array}
\usepackage{slashed}
\usepackage{afterpage}
\usepackage{cancel}
\usepackage{appendix}
\usepackage{multirow}
\usepackage{feynmp}
\newcommand*\mvl[2]{\rlap{\small\kern-#1em\relax#2}}

\newcommand{\MET}{\cancel{E}_T}
\newcommand{\MDM}{m_\chi}
\newcommand{\Oq}{\mathcal{O}_q }
\newcommand{\Og}{\mathcal{O}_g }

\title{Identifying dark matter interactions in monojet searches}
\author[a]{Prateek Agrawal}
\affiliation[a]{
Fermilab National Accelerator Laboratory,\\
Batavia, IL, 60510, USA
}

\author[b]{and Vikram Rentala}
\affiliation[b]{
Department of Physics \& Astronomy,\\
Michigan State University, E. Lansing, MI 48824, USA}

\emailAdd{prateek@fnal.gov}
\emailAdd{rentala@pa.msu.edu}

\abstract { We study the discrimination of quark-initiated jets from
gluon-initiated jets in monojet searches for dark matter using the
technique of averaged jet energy profiles.  We demonstrate our results
in the context of effective field theories of dark matter interactions
with quarks and gluons, but our methods apply more generally to a wide
class of models.  Different effective theories of dark matter and the
standard model backgrounds each have a characteristic quark/gluon
fraction for the leading jet.  When used in conjunction with the
traditional cut-and-count monojet search, the jet energy profile can
be used to set stronger bounds on contact interactions of dark matter.
In the event of a discovery of a monojet excess at the 14 TeV LHC,
contact interactions between dark matter with quarks or with gluons
can be differentiated at the 95\% confidence level.  For a given rate
at the LHC, signal predictions at direct detection experiments for
different dark matter interactions can span five orders of magnitude.
The ability to identify these interactions allows us to make a tighter
connection between LHC searches and direct detection experiments.  }

\preprint{FERMILAB-PUB-13-556-T, MSU-16122013}

\begin{document}
\maketitle

\section{Introduction}
\label{sec:intro}
There is overwhelming evidence for the existence of dark matter (DM)
from astrophysical and cosmological observations. While its
gravitational interactions are well understood, its particle
properties such as its mass and other interactions have so far eluded
detection.

There are a variety of experiments which are searching for DM
particles (see \cite{Bertone:2004pz} for a review). Direct detection
experiments are looking for nuclear recoils from scattering with the DM in our immediate vicinity
~\cite{
Benetti:2007cd,
Ahmed:2009zw,
Bernabei:2010mq,
Akimov:2011tj,
Angloher:2011uu,
Armengaud:2012pfa,
Aalseth:2012if,
Kim:2012rza,
Behnke:2012ys,
Baudis:2012zs,
Agnese:2013rvf, Akerib:2013tjd}.
Indirect detection involves searching for annihilation products of DM
from regions of high DM density~\cite{
Barwick:1997ig,
Desai:2004pq,
Aguilar:2007yf,
Chang:2008aa,
FermiLAT:2011ab,
Aartsen:2012kia,
fortheFermiLAT:2013naa,
Adriani:2013uda,
Fermi-LAT:2013uma,
Ackermann:2013yva}.
 In this paper we will focus on collider searches for DM~\cite{
Abbiendi:2000hh,
Heister:2002ut,
Abazov:2003gp,
Abdallah:2003np,
Achard:2003tx,
Abdallah:2008aa,
Aad:2011xw,
Aaltonen:2012ek,
Aaltonen:2012jb,
ATLAS:2012ky,
Chatrchyan:2012me,
Chatrchyan:2012tea,
Aad:2013oja}. At the
Large Hadron Collider (LHC), DM can be pair produced at a relatively
high rate through interactions with colored particles in the standard
model (SM). Associated production of initial state radiation (ISR)
jets can then lead to a
signature with energetic jets with significant missing transverse
energy ($\MET$), which has been called the ``monojet'' signal.

In the limit of a heavy mediator for interactions between the DM
particles and the SM, we can use an effective field theory (EFT)
description with contact interactions involving the DM and SM quarks
and gluons to study the expected monojet
signal~\cite{Goodman:2010yf,
Goodman:2010ku,Bai:2010hh,
Rajaraman:2011wf,Fox:2011pm}.
At the LHC, the interaction is probed at high energies, and the
validity of the EFT can be circumspect. One way to derive a more
robust bound is to include the mediator in a simplified model
framework~\cite{Buchmueller:2013dya}. In addition to the monojet
signature, there may also be related signatures of other visible
particles produced in association with the DM \cite{Zhou:2013fla},
such as mono-photons \cite{Gershtein:2008bf}, mono-W/Z
\cite{Bai:2012xg,Bell:2012rg,Carpenter:2012rg} and mono-Higgs
\cite{Petrov:2013nia, Carpenter:2013xra}.

There are three main goals for monojet searches that can be phrased in
the EFT framework:
\begin{enumerate}
  \item In the absence of discovery, to set a limit on the
    suppression scale of new physics operators that couple DM to SM
    particles.
  \item Upon a discovery of an excess at the LHC,
    to try and identify the new physics operator or
    combinations of operators that are responsible for the signal.
  \item To verify whether the particles escaping the detector and
    reconstructed as missing transverse momentum in monojet searches
    constitute the DM of the universe.
\end{enumerate}

In this paper, we will consider two cases: one where DM has contact
interactions with quarks and the other where it has contact
interactions with gluons. Our main idea is to exploit the fact that
\textsl{these two scenarios for DM interaction and the SM background,
each have a characteristic quark/gluon fraction associated with the
parton that initiates the monojet}.  Therefore, any technique which
distinguishes between quark- and gluon-initiated jets will effectively
discriminate between these DM production operators and also the SM
background. In this paper we will use averaged jet energy profiles for
such a discrimination.  This method can be used to both improve the
sensitivity of monojet searches or to identify the type of contact
interaction in the event of a discovery of a monojet signal.

The EFT approach simplifies the connection between predictions at the
LHC and direct detection experiments.
In direct detection experiments, since the DM-nucleon scattering is probed at low
momentum transfer, the EFT can be used to reliably calculate the expected
signal. However, given a certain
monojet cross section at the LHC, different interactions lead to
direct detection cross sections that span five orders of
magnitude.  Therefore, the ability to discriminate between DM
interactions with quarks versus interactions with gluons would have
major ramifications for direct detection experiments.

For low DM masses ($\lesssim 10$ GeV), the current experimental limits
from the ATLAS and CMS collaborations constrain the suppression scale
of the contact interactions much more strongly than current
spin-independent direct
detection bounds \cite{ATLAS-CONF-2012-147,CMS-PAS-EXO-12-048}.  In
the event of a monojet signal in this region, this technique could be
used to identify the target search region for future experiments.  For
higher DM masses, strong limits have been set in direct detection
experiments. Thus, given a monojet signal in this region, this
technique could help us to evaluate the hypothesis that the particles
produced in the monojet signal are the same as the galactic DM.

This paper is organized as follows. In Sec.~\ref{sec:MJ}, we review the
current monojet searches at LHC in terms of EFT operators and the
current bounds on their suppression scales. In Sec.~\ref{sec:qgf}, we
will point out how using the quark/gluon fraction of the monojet can
improve the bounds or help discriminate between DM contact
interactions with quarks versus gluons. Then,
in Sec.~\ref{sec:JEPs}, we will review the concept of jet energy
profiles (JEPs) to distinguish between quark and gluon jets. In
Sec.~\ref{sec:methods}, we will describe our methodology for applying
the JEPs to monojet searches. In Sec.~\ref{sec:results} we will
present our results for improving the limits on the suppression scale
of the EFT operators using JEPs at the LHC.
We will also show how well the two types of DM operators can be distinguished
at the 14~TeV LHC assuming discovery of an excess in monojets. In
Sec.~\ref{sec:directdet}, we will discuss the implications for direct
detection searches for dark matter. Finally, we will conclude in
Sec.~\ref{sec:concl}.

\section{Dark matter searches with monojets}
\label{sec:MJ}
In this section we discuss the EFT approach to study DM contact interactions with quarks and gluons. We will summarize the current limits placed on the scale of the contact interactions by monojet searches at the LHC.

\subsection{The effective field theory approach}
The EFT approach to DM monojets at the LHC was first studied
in~\cite{Goodman:2010yf,Goodman:2010ku,Bai:2010hh,Rajaraman:2011wf,Fox:2011pm}.
We will adopt the same approach as in these papers. We will consider
Dirac fermion DM with contact operators of dimensions 6 and 7 that
couple pairs of dark matter particles to SM light quarks and gluons
respectively.

For simplicity, we will will focus on just two operators,
\begin{align}
\Oq = \frac{1}{M^2_*} (\bar{q}  \gamma_\mu q)( \bar{\chi} \gamma^\mu \chi), \\
\Og = \frac{\alpha_s}{4 M^3_*} (G^a_{\mu\nu} G^{a\mu\nu}) (\bar{\chi} \chi).
\end{align}
Here, the normalization has been chosen in accord with
Ref.~\cite{Goodman:2010ku}, where $\mathcal{O}_q$ and $\mathcal{O}_g$
are referred to as D5 and D11 in their study. We will henceforth refer to these models as QDM and GDM to indicate the coupling of dark matter to quarks and gluons respectively. Each of these models can be parameterized with only two parameters: the dark matter mass, $\MDM$ and the suppression scale, $M_*$.

Some alternate choices for the quark contact operators, which we do not consider, involve different bilinears for the quarks. The scalar quark bilinears $(\bar{q}_L q_R)$, $(\bar{q}_R q_L)$ are
expected to be suppressed by the light quark masses and thus the monojet search would give rise to very weak limits on these models. The vector $(\bar{q} \gamma_\mu q)$ and axial vector
$(\bar{q} \gamma_\mu \gamma_5 q)$ bilinears give practically similar
limits at collider searches, but the difference is important when we
consider direct detection experiments because the former gives rise to
spin-independent interactions with nuclei, which are strongly
constrained, whereas the latter give spin-dependent interactions,
which are weakly constrained.

\subsection{Searches and limits at the LHC}
We now discuss how monojet searches at the LHC constrain the parameter space of the EFTs for DM.
The operators $\Oq$ and $\Og$, combined with ISR, can give rise to monojet
signatures defined as a high energy jet accompanied by large $\MET$ at
the LHC. The most dominant irreducible SM background comes from $Z + j$ with $Z
\rightarrow \nu \bar{\nu}$. There is also a sub-dominant irreducible background from $W
+ j$ with $W \rightarrow \tau \nu$,  where the $\tau$ decays
hadronically. In addition there are reducible backgrounds from $W
+ j$ with $W \rightarrow \mu \nu$ or $e \nu$, where the lepton
is either missed or misidentified as a jet. We will refer to all these sources of SM background as the $V+j$ background.


In order to discover an excess in monojets a \textsl{cut-and-count}
strategy is used by the ATLAS and CMS experiments
\cite{ATLAS-CONF-2012-147, CMS-PAS-EXO-12-048}. The typical $p_T$ spectrum of the monojet
associated with DM production is more slowly falling relative to the
$p_T$ spectrum of the background. By choosing different values of the $p_T$ and
$\MET$ cut, the experimental analysis defines different ``signal
regions'' (SR) in which limits are placed on the suppression
scale, $M_*$ for a given value of $\MDM$. Choosing too low
a value of the $p_T$ and $\MET$ cuts leads to large systematic errors on the predicted number of background events, whereas too high a value of the $p_T$
cut reduces the number of events on which to perform statistical analysis.

We will adopt the so-called SR3 region from the ATLAS search, which
they use to set limits on the scale $M_*$. In this region, the
statistical errors are comparable to the systematic errors.

The following cuts are required for event selection in this region:
\begin{itemize}
\item $\MET > 350$~GeV,
\item leading jet with $p_T > 350$~GeV and $|\eta| < 2.0$,
\item at most two jets with $p_T > 30$~GeV and $|\eta| < 4.5$,
\item $\Delta \phi$ (jet, $\MET$) $> 0.5$ (second leading jet),
\item lepton vetoes.
\end{itemize}

\begin{figure}
\centering
\begin{subfigure}{.5\textwidth}
  \centering
  \includegraphics[width=7.5cm,keepaspectratio]{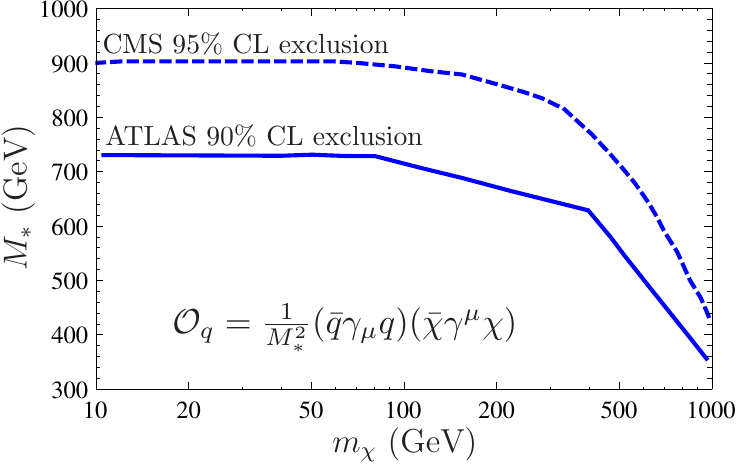}
  \caption{Limit on $M_*$ in the QDM model.}
  \label{fig:sub1}
\end{subfigure}%
\begin{subfigure}{.5\textwidth}
  \centering
 \includegraphics[width=7.5cm,keepaspectratio]{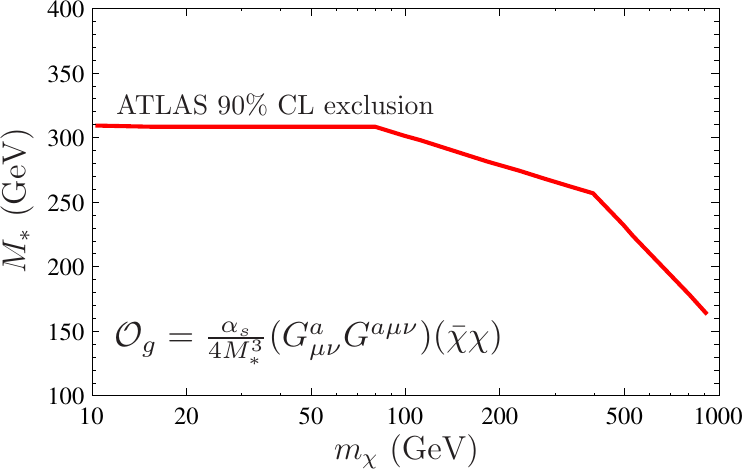}
  \caption{Limit on $M_*$ in the GDM model.}
  \label{fig:sub2}
\end{subfigure}
\caption{ATLAS \cite{ATLAS-CONF-2012-147} and CMS \cite{CMS-PAS-EXO-12-048} exclusions on the suppression scale $M_*$ for various dark matter masses from the 8 TeV run of the LHC. ATLAS used 10.5~fb$^{-1}$ of luminosity for their analysis, whereas CMS used 19.5~fb$^{-1}$ of luminosity. CMS has not yet released an official limit for the $O_g$ operator.}
\label{fig:explimits}
\end{figure}

The resulting limits in the $\MDM - M_*$ plane from non-observation of an excess in monojets at the 8~TeV run of the LHC are shown for the operators $\Oq$ and $\Og$ in Fig.~\ref{fig:explimits}. Strictly speaking, the validity of the effective theory description restricts the values of $E \lesssim M_*$, where $E$ is the typical energy scale at the monojet vertex. However, even for regions where the effective theory is invalid, the experimental bounds in the $\MDM - M_*$ plane can give a qualitative understanding of the bounds on dark matter physics models.

\begin{figure}
\centering
\begin{subfigure}{.5\textwidth}
  \centering
  \includegraphics[width=7.cm,keepaspectratio]{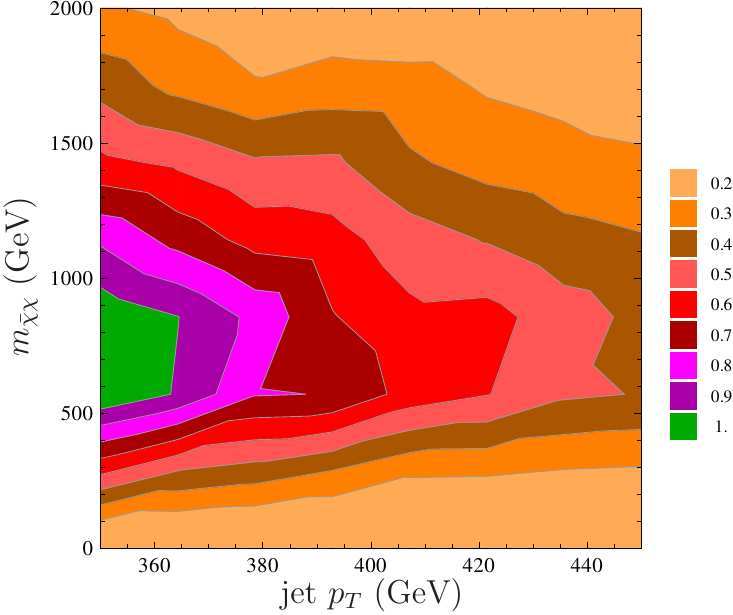}
  \caption{Event kinematic distribution for $\Oq$.}
  \label{fig:sub3}
\end{subfigure}%
\begin{subfigure}{.5\textwidth}
  \centering
  \includegraphics[width=7.cm,keepaspectratio]{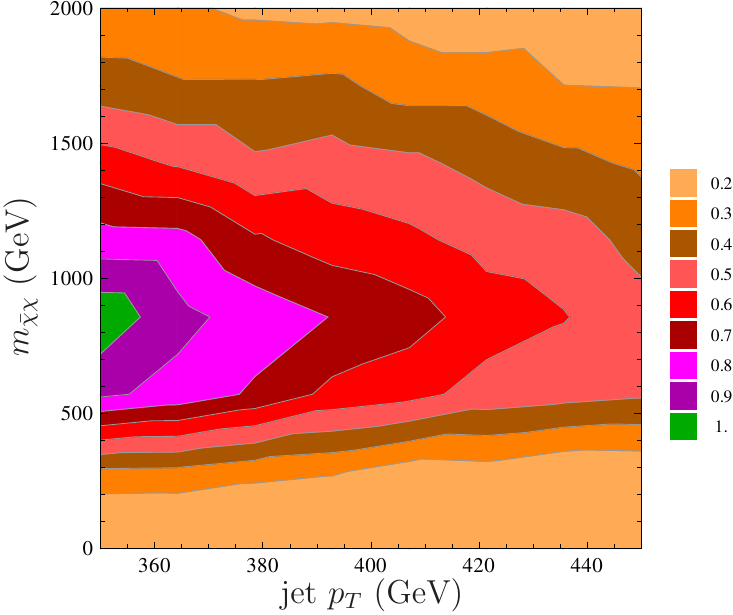}
 \caption{Event kinematic distribution for $\Og$.}
  \label{fig:sub4}
\end{subfigure}
\caption{Plot of the distribution of monojet signal events, at the 8 TeV LHC, as a function of invariant mass of the $\bar{\chi}\chi$ and jet $p_T$ in different EFTs. The normalization of the event distribution is arbitrary. We can see that most events are produced with a large center of mass energy, close to the current bound on the scale of the effective operators, $M_*$ .}
\label{fig:kinematics}
\end{figure}

To gain an understanding of the typical kinematics of the monojet signal, we plot the expected event distribution as a function of the invariant mass of the DM particles ($m_{\bar{\chi}\chi}$) and the jet $p_T$ in Fig.~\ref{fig:kinematics}. Practically of course, only the jet $p_T$ of the events is observable, however, this plot is instructive because we can see that the relevant typical energy scale, $E \simeq \sqrt{m^2_{\bar{\chi}\chi} + p^2_T}$, is close to the current bounds on the effective theory scale $M_*$. Nevertheless, we will continue to use the EFT description and it will not affect our main conclusions.

\section{Quark/Gluon composition of the monojet}
\label{sec:qgf}
\begin{figure}
\centering
\begin{subfigure}{.5\textwidth}
  \centering
  \includegraphics[width=7.5cm,keepaspectratio]{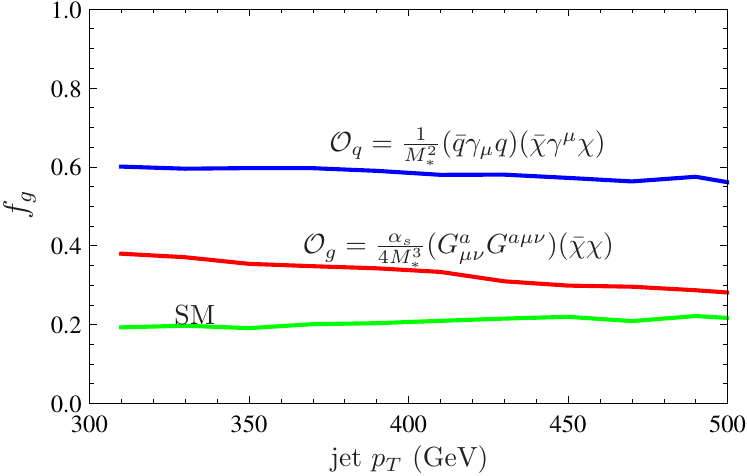}
  \caption{Gluon fraction vs $p_T$.}
  \label{fig:gfractionsa}
\end{subfigure}%
\begin{subfigure}{.5\textwidth}
  \centering
  \includegraphics[width=7.5cm,keepaspectratio]{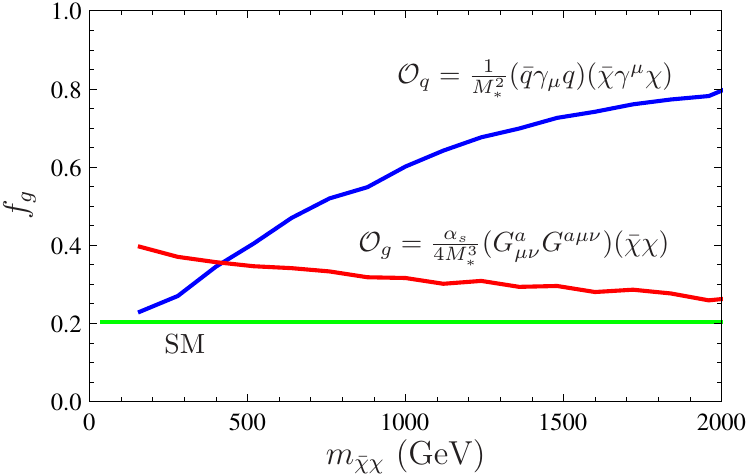}
  \caption{Gluon fraction vs $m_{\bar{\chi}{\chi}}$.}
  \label{fig:gfractionsb}
\end{subfigure}
\caption{Plots of the gluon fractions $(f_g)$ at the 8 TeV LHC as (a) a function of jet $p_T$ for the monojet signal and the SM background and (b) $m_{\bar{\chi}{\chi}}$ for the monojet signal. In (b) the SM $(f_g)$ value is shown only for reference.  The differing gluon fractions of the signals and the background as a function of the jet $p_T$ suggest that experimental separation based on this variable could allow for better separation of signal and background than a counting experiment alone. In the event of a discovery this technique could also allow us to identify the operator that is responsible for the signal.}
\label{fig:gfractions}
\end{figure}

It is interesting to consider the composition of quarks/gluons in monojet signals and backgrounds.
This can be parameterized as a gluon fraction $f_g$. The typical gluon fractions for a sample of monojets generated
from a pure signal sample from $\Oq$, $\Og$ and a pure background
sample from $V + j$ are shown as a function of the jet
$p_T$ in Fig.~\ref{fig:gfractionsa} and as a function of
$m_{\bar{\chi}{\chi}}$ in Fig.~\ref{fig:gfractionsb}.

We note some interesting features of the gluon fractions,
\begin{itemize}
\item The QDM sample at all jet $p_T$s is slightly richer in gluons than in quarks $(f_g\simeq
  0.6)$, while the $V + j$ sample is quark rich $(f_g\simeq 0.2)$.
  The $V+j$ and QDM processes are generated by similar processes with similar diagrams, so this result seems surprising at first. We shall explain this discrepancy in the gluon fractions below.

  At the parton level, there are two main processes that contribute to $V/(\bar{\chi}{\chi})+j$ are $q \bar{q} \rightarrow V/(\bar{\chi}{\chi}) g$ and $q g \rightarrow V/(\bar{\chi}{\chi}) q$. For fixed energies for the initial partons, in both the SM and QDM monojets, the rate for the $q \bar{q}$ initiated process typically dominates the $q g$ initiated process by a factor of a few.

  However, the main distinction is that the QDM kinematics is controlled by $m_{\bar{\chi}{\chi}}$ (Fig.~\ref{fig:kinematics}) and is probed at high momentum fraction ($x$) for the partons, whereas the $V+j$ kinematics is controlled by the mass of the W and Z bosons ($m_W/m_Z$) and is probed at a relatively low value of x. This leads to the $qg$ initiated process being highly enhanced for $V+j$ due to the large gluon parton density function (pdf), which leads to the SM monojet being quark rich. However, the QDM monojet is produced at larger $x$ and the enhancement of the gluon pdf relative to the sea $\bar{q}$ pdf is smaller and both processes contribute similarly leading to roughly similar quark and gluon fractions. We can see this effect in Fig.~\ref{fig:gfractionsb}, where the gluon fraction for the QDM
  signal starts of the same as the SM when $m_{\bar{\chi}{\chi}} \simeq m_W/m_Z$ (low $x$ values) and rises strongly with the mass of the $\bar{\chi}\chi$ (high $x$ values).

\item The GDM sample starts off at $(f_g\simeq 0.4)$ near the $p_T$
  cut threshold and drops to $(f_g\simeq 0.2)$ for higher $p_T$s. The
  GDM production takes place through the processes $gg \rightarrow
  \bar{\chi}{\chi} g $ and $gq \rightarrow \bar{\chi}{\chi} q$.  Here
  the $gq$ initiated process dominates for fixed partonic energies.
  However, at low values of $p_T$ or $m_{\bar{\chi}\chi}$ the gluon
  pdfs lead to the $gg$ process being enhanced and hence $f_g \simeq
  0.4$. At large values of $p_T$ or
  $m_{\bar{\chi}\chi}$ the $gq$ process dominates due to the valence
  quark pdf being probed at high $x$ and we see that the monojet
  sample becomes quark rich ($f_g\simeq 0.2$).
\end{itemize}

The characteristic distinction between $f_g$ in each of the three
cases strongly suggests that identification of the gluon fraction of
the monojet sample is a good way to enhance sensitivity in monojet
searches by providing an additional handle to discriminate signal from
background. The distinction can also be used to discriminate between
different signal hypotheses once an excess in monojets is discovered.
We will study the use of jet energy profiles (JEPs) in the next
section to see how the gluon fractions can be probed.

\section{Jet energy profiles}
\label{sec:JEPs}
\begin{figure}
\centering
\begin{subfigure}{.5\textwidth}
  \centering
  \includegraphics[width=7.5cm,keepaspectratio]{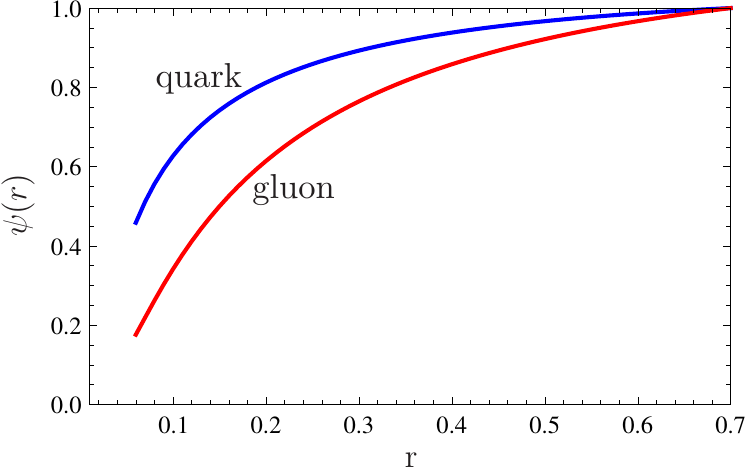}
  \caption{}
  \label{fig:qgjep}
\end{subfigure}%
\begin{subfigure}{.5\textwidth}
  \centering
  \includegraphics[width=7.5cm,keepaspectratio]{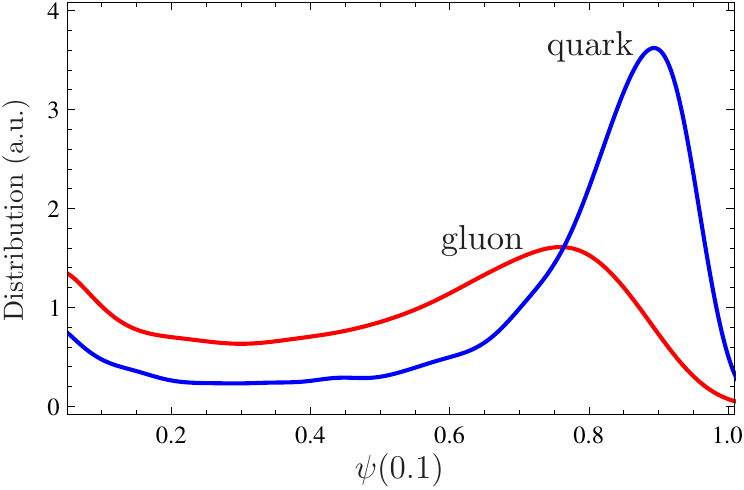}
  \caption{}
  \label{fig:qgoneevt}
\end{subfigure}
\caption{(a) Typical quark and gluon jet energy profiles, $\psi(r)$. Quarks typically radiate less and have a more steeply rising JEP as compared to gluons. (b) Distribution (with arbitrary normalization) of $\psi(0.1)$ for quark and gluon jets generated in PYTHIA 8 \cite{Sjostrand:2007gs}. Individual jets may fluctuate in their profiles significantly. We can see here the Sudakov tail of the radiation pattern.}
\end{figure}

We now describe the technique of using jet energy profiles (JEPs) to identify the gluon fraction of a jet sample.
We will consider anti-$k_T$ jets \cite{Cacciari:2008gp} with a cone size $R = 0.7$ as measured in the rapidity-azimuth $(\eta-\phi)$ plane. The JEP, $\psi(r)$, for an individual jet is defined as the fraction of jet $p_T$ inside a subcone of size $r$ relative to the total jet $p_T$,
\begin{equation}
\label{eq:psi}
\psi(r) = \frac{ \int\limits^{r}_{0}{\frac{dp_T}{dr'} dr'}}{\int\limits^{R}_{0}{\frac{dp_T}{dr'} dr'}},
\end{equation}
where the denominator is just the total $p_T$ of the jet. The JEP is defined so that $\psi(0) = 0 $ and $\psi(R) = 1$. Gluon initiated jets are expected to spread more due to more radiation and thus have a slowly rising JEP. Quark initiated jets on the other hand radiate less, and thus accumulate a larger fraction of their energy for fairly small $r$ and have a quickly rising JEP. A typical quark/gluon profile is shown in Fig.~\ref{fig:qgjep}.

For an individual quark/gluon the JEP fluctuates widely due to a Sudakov tail as shown in Fig.~\ref{fig:qgoneevt}. The characteristic shape and spread of the JEP also depends on the jet $p_T$. This variance makes an individual quark/gluon jet difficult to tag. However, on average, a sample of pure quark or gluon jets can be distinguished fairly well statistically as we shall see.

\subsection{Determining the JEPs of a sample of pure quarks/gluons}
There are several different approaches that one could take towards determining the average JEP of a sample of pure quark or gluon events.
\begin{enumerate}
\item Experimental: data driven approach,
\item Theory: perturbative QCD (pQCD) calculation,
\item Simulation: Monte-Carlo shower and hadronization generators.
\end{enumerate}

The best approach to determining the JEPs is to use all three techniques in conjunction. We will discuss the merits and demerits of each approach below.

\subsubsection{Experimental JEPs}
JEPs have been measured at the Tevatron CDF experiment \cite{Acosta:2005ix} and at the LHC by both ATLAS \cite{Aad:2011kq} and CMS \cite{CMS-PAS-QCD-10-013}. Assuming factorizability (i.e. universal behavior) of quark/gluon jets of a given $p_T$, we could measure the average quark/gluon profile in control samples of pure (or theoretically well known fractions) of quark/gluon jets. The experimental JEPs have the advantage of having very small error bars and thus allow for a very precise determination of the average quark/gluon jet profile for a given $p_T$. However, factorizability must be proven theoretically to guarantee the
validity of using calibrated quark/gluon JEPs in a control sample to
match the behavior of quark/gluon JEPs initiated from a different hard
process. Theoretical calculations may also be needed to extrapolate the behavior of JEPs
with $p_T$s outside of the experimental control regions.

\subsubsection{Perturbative theory calculation}
The JEPs can be calculated using the techniques of pQCD. At
next-to-leading order (NLO) the profiles were found to overshoot the data
\cite{Li:2011hy} at small $r$, where the logarithmic corrections
$\alpha_s\log(R/r)$ are important and need to be resummed. The fixed
order results are thus inappropriate for comparison with experiment.
The next-to-leading-logarithm (NLL) terms of the form $\alpha^n_s
\left (\log(R/r) \right )^{2n}$ and $\alpha^n_s \left (\log(R/r)
\right )^{2n-1}$  need to be resummed to all orders in $\alpha_s$ to
obtain a finite answer. The resummed JEPs were calculated in
\cite{Li:2011hy,Li:2012bw}. Very good agreement was found between the
experimental data and the resummation calculation for a wide range of
jet $p_T$s. It was also theoretically confirmed that a gluon jet is
broader than a quark jet with the same $p_T$. The NLL resummed JEPs
capture the behavior of the experimental JEPs extremely well and
demonstrate factorizability.

In order to match the theoretical prediction to the experimental
results, it is necessary to fix some (arbitrary) scale parameters
which are introduced into the pQCD calculations to estimate the
effects of the sub-leading logarithmic contributions. Varying the
scale parameters allows us to estimate the error on the theoretical
calculation. The theoretical error on the JEP is large compared to the experimental
uncertainties on JEPs. However, upon calibrating to the data in a
known control region, the scale parameters can be fixed and the
resulting JEPs can be extrapolated outside of the control regions of
experiments.

The resummed theory calculation only predicts the average JEP for a
sample of quarks and gluons. By itself, it does not show the
fluctuations that would be needed to determine the statistical
uncertainty of the JEP at an experiment.


\subsubsection{Monte-Carlo simulations}
Monte-Carlo shower and hadronization generators such as PYTHIA
\cite{Sjostrand:2007gs} can also be used to produce the JEPs. The
predictions of the Monte-Carlo are notoriously tune-dependent and need
to be calibrated to match the data, for e.g. PYTHIA tune A
\cite{PYTHIAA} can describe the observed JEPs well.  The advantage of
the Monte-Carlo method is that it allows us to generate event samples
that can be used to predict the statistical uncertainty of the JEP.
\bigskip

As theorists we have access only to the latter two methods. In order
to calculate the central value of the average JEPs, we will use the results of
the NLL resummation calculation. We fix the values of the integration
constants by matching the theory calculation to Tevatron data. In
order to estimate the statistical uncertainty, we will use the default
tune of PYTHIA 8 to generate fluctuations and transpose the error bars
obtained in this manner onto the pQCD JEPs.

\subsection{From individual quark/gluon JEPs to weighted samples}
As noted above, there are large Sudakov tails in the JEP distribution for individual quarks/gluons.
In order to reduce sensitivity to the shape of this distribution, we will
work with the profile averaged over a sample of jets. The immediate advantage is
that the \textsl{average profile} is Gaussian distributed if the sample size is
sufficiently large, due to the central limit theorem. The error bars
on the profiles then scale simply with the size of the sample ($N$) as
$1/\sqrt{N}$.

It is also convenient to work at the level of individual models, instead of
a pure quark or gluon sample. The pQCD calculation provides the
expected average JEP for pure samples of quarks or gluons in a
given $p_T$ bin. Thus, given the quark/gluon composition in respective
$p_T$ bins from the model, we can predict the average profile for a sample.
The average profile is a weighted average of quark and gluon JEPs, with the weight depending on
the underlying process, parton distribution functions, collider energy and jet $p_T$.
For the processes we consider, the profile has a very mild dependence on the $p_T$ of the jet, due to both $p_T$-dependent QCD effects as well as
different quark-gluon fractions in different $p_T$ bins (see Fig.~\ref{fig:gfractionsa}).

Averaging over $p_T$ bins, we arrive at a single profile which we call the ``expected average profile'' (EAP),
\begin{equation}
\label{eq:psiavg}
\psi(r)_\textrm{EAP} = \int \left ( \frac{dN_q}{dp_T}\psi_q(r,p_T) + \frac{dN_g}{dp_T}\psi_g(r,p_T) \right) dp_T / (N_q + N_g).
\end{equation}
Here, $N_q$ and $N_g$ are the number of quarks and gluons initiating the jets in the sample. $\psi_q(r, p_T)$ and $\psi_g(r, p_T)$ are the \textsl{average quark and gluon profiles} for jets of a particular $p_T$, which can be calculated from the pQCD resummation technique. The statistical fluctuations in the average JEP are controlled by \textsl{both} the statistical fluctuations of the gluon fraction of the sample \textsl{and} the statistical fluctuations of the profiles of individual quark and gluons jets (Fig.~\ref{fig:qgoneevt}). The size of these statistical fluctuations can be determined by Monte-Carlo methods which simulate the hard process (which calculates the fluctuation in the quark/gluon fractions) and the soft/collinear processes (which leads to fluctuations in the individual quark/gluon JEPs).


\section{Methodology}
\label{sec:methods}
We now discuss the application of JEPs to monojet searches.
As discussed, we will consider two EFTs for dark matter, labeled
QDM and GDM,
corresponding to the following interactions respectively
\begin{align}
\Oq = \frac{1}{M^2_*} (\bar{q}  \gamma_\mu q)( \bar{\chi} \gamma^\mu \chi), \\
\Og = \frac{\alpha_s}{4 M^3_*} (G^a_{\mu\nu} G^{a\mu\nu}) (\bar{\chi} \chi).
\end{align}

In Sec.~\ref{sec:qgf}, we saw that the QDM and GDM models and the SM
background each have characteristic quark/gluon fractions. In this section
we will separate the different DM models and
the SM background using JEPs. We will benchmark against the 10.5
fb$^{-1}$ ATLAS results \cite{ATLAS-CONF-2012-147}, since their
analysis sets limits for both the QDM and GDM EFTs that we are
interested in.

The QDM and GDM models are each characterized by two parameters, the dark
matter mass, $\MDM$ and the suppression scale of the effective operator,
$M_*$. The dark matter mass affects the production cross section, as
well as the $p_T$ spectrum of the radiated jets in monojet searches.
We simulate samples for dark matter masses in the range $10$ to $1000$
GeV. Within the region of validity of the EFT, the suppression scale $M_*$
only affects the total rate of production. Therefore, predictions for
varying values of $M_*$ can be easily obtained by scaling the expected
number of events.

For each signal + background model, we have two observables to calculate: the expected number of events (N) and the EAP of the monojet $\psi(r)_\textrm{EAP}$. These observables can then be compared to the corresponding expectations from the background only hypothesis.

We will evaluate the advantage of including the JEP over just a cut-and-count experiment in two contexts:
\begin{enumerate}
 \item Improvement of the
exclusion limits on $M_*$ as a function of $\MDM$ for 8 TeV monojet
searches.
\item Higher discovery reach, and the power to distinguish the QDM and GDM EFTs at the 14 TeV LHC.
\end{enumerate}

In order to determine our ability to separate $N$ and $\psi(r)_\textrm{EAP}$ under different hypotheses, we need to estimate the errors on these quantities.
Below, we will discuss how we obtain the statistical errors on these quantities by conducting pseudo-experiments.

We simulate monojet samples for the QDM + SM, GDM + SM and SM only scenarios at the parton level
with MadGraph5 \cite{Alwall:2011uj} for a range of dark matter masses.
We then shower the events using PYTHIA and use SpartyJet
\cite{Delsart:2012jm} (a
wrapper for FastJet \cite{Cacciari:2011ma}) to perform the jet clustering using the
anti-$k_T$ algorithm. The cuts on the event sample for the SR3 region
(see Sec.~\ref{sec:MJ}) are then applied to these clustered events.
For a given luminosity we can now predict the expected number of
events $N_\textrm{exp}$ and its statistical fluctuations
($\sqrt{N_\textrm{exp}}$) in each model.

We then consider the leading $p_T$ jet for each event in the sample.
We construct the jet energy profile $\psi(r)$ of the leading jet by
sampling at seven different values of $r = 0.1 \ldots 0.7$, as would be
expected in a realistic calorimeter simulation. From this collection
of JEPs we calculate the average JEP. By running many
pseudo-experiments we can also predict the statistical fluctuation on
$\psi(r)$ for each value of $r$.

The average JEP for a large number of events should in principle
reproduce the EAP. As mentioned in the previous section, this
determination is subject to large uncertainties from the PYTHIA tune.
We will therefore use the theory prediction to derive the EAP in each
model by constructing it from the pQCD calculation and the leading
order prediction of the quark/gluon fractions at different $p_T$s from
MadGraph5. We will, however, retain the statistical error bars from
PYTHIA on the JEPs.

Before moving on to our results, we present a few important details
for our analysis in this section.

\subsection{Background}
\label{subsec:bkg_methods}
The dominant SM background arises from $Z/W +j$, with the $Z$
decaying to neutrinos, and the $W$ decaying to hadronic taus, or to
leptons which are either missed by the detector or misidentified as
jets. This latter component of the background is sensitive to detector
effects which are hard to model. However, the gluon fraction of the associated jet is nearly identical for
both the $Z+j$ and the $W+j$ background. Therefore, we
choose to model the entire background as a scaled-up $Z+j$ background.
For the 8 TeV analysis, we simulate the background by generating the $Z+j$ background with all the relevant cuts outlined in Sec.~\ref{sec:MJ},
and then scale the simulated cross section to agree with the expected number of events for the 10.5 fb$^{-1}$ ATLAS study \cite{ATLAS-CONF-2012-147}. This scaling also includes the k-factor for the full $V+j$ background. For 14 TeV projections, we simulate $Z+j$ at the 14 TeV LHC, and apply the
same scaling factor as in the 8 TeV case, to get the total background
prediction.

The really crucial ingredient for background modeling comes from the systematic
uncertainties. We discuss this in detail in the next subsection.

\subsection{Systematic uncertainties}
We will now discuss the systematic uncertainties on each observable in
our analysis -- the total number of events, and the
jet energy profile.

The systematic error on the total number of events is controlled by systematics on the SM background rate as well as on the QDM/GDM signal rate.
For the 8 TeV ATLAS analysis \cite{ATLAS-CONF-2012-147}, the background was estimated from a
combination of data driven and Monte Carlo simulations. Therefore, the
systematic uncertainty on the background has contribution from the control region
statistical uncertainty, the simulation statistical uncertainty, and
uncertainty in the jet energy scale and other uncertainties, leading
to a $7\%$ uncertainty on the background rate in the relevant signal region. For our 8~TeV
exclusion analysis, we will add this systematic error in quadrature with our statistical errors when modeling the fluctuations of the background.  The
corresponding systematic uncertainties at 14~TeV needs a detailed
study. We will choose $5\%$ systematic on the background rate as a benchmark at 14~TeV.

We do not include systematic uncertainties for the signal. While the
number of signal events itself is subject to systematic uncertainties,
these are traditionally accounted for by presenting exclusions for
$\pm1\sigma$ systematic variation of the predicted cross section. In
our study we only present the limits for the nominal signal
prediction.

The systematic uncertainty on our modeling of the EAP is important
to consider. By examining our formula for the EAP, Eq. (\ref{eq:psiavg}), we see that there are several sources of systematic uncertainty.
Firstly, we need to know the gluon fraction as a function of the jet $p_T$. We are only using the leading order prediction for $f_g(p_T)$, but NLO predictions for both the signal and background are needed to make a precise prediction on this fraction. We will neglect this systematic because typically the ratios of rates have a smaller uncertainty than the total rate itself. This is a relevant issue and warrants further study in future work.
Secondly, the average JEP for a pure quark or gluon sample is determined from the pQCD NLL calculation calibrated to data to fix the integration constants. In this kind of data driven determination, the statistical uncertainty of the control sample will contribute, but this error as mentioned earlier is extremely small and can be ignored. However, outside of the control $p_T$ regions, the errors from the theory calculation can contribute to the systematic uncertainty on the JEP.
Finally, from experimental considerations, the measured JEP at the 14 TeV LHC will be subject to increased pile-up activity which implies that there would be much more
extraneous radiation which could potentially get clustered with jets arising from the hard process, which will tend to smear the
distinction between quark and gluon jets (see \cite{Gallicchio:2012ez} for one strategy for dealing with pile-up). This source of systematic needs to be studied in detail once the ATLAS and CMS experiments start collecting data at 14 TeV.

In summary, we will neglect all systematic contributions to the uncertainty on the EAPs for the pure GDM, pure QDM and pure SM samples, and we will only estimate the statistical errors on these EAPs.

\subsection{Projecting from JEPs to the signal/background fraction}
For each hypothesis (QDM + SM, GDM + SM and SM only), we can determine the average JEP for the monojet sample and its statistical fluctuations.
We would like to project this average JEP onto a single variable that can be compared for these different models.

We define a one-variable parametrization $f$ of the profile, which
interpolates between a SM EAP ($f=0$) and a QDM
EAP ($f=1$). For a given sample, $f$ is defined as the best-fit
parameter of the average profile to the following expression,
\begin{align}
\psi_f(r)
&=
f \psi^{\textrm{QDM}}_\textrm{EAP}(r) +(1-f) \psi^{\textrm{SM}}_{\textrm{EAP}}(r)
\end{align}
For a sample of QDM + SM events $f$ can be approximately understood as the fraction of signal events in the sample averaged over all $p_T$ bands.
It is worth noting that even for the GDM model, we
evaluate $f$ as interpolating between the QDM and the SM
profile. This is due to the fact that the
gluon fraction for the GDM sample resembles more closely the
gluon fraction of the SM sample, and hence the profiles are not well
separated from each other.

The fluctuations in the pseudo-experiment average JEPs can then be translated into fluctuations in $f$. The expected value of $f$ will also have systematic uncertainties due to the systematic errors on the JEP (which we neglect) and systematic uncertainty in the prediction of the number of background events (which we include).

\bigskip

We have now parameterized every pseudo-experiment monojet sample by two variables, its JEP
projected on to the parameter $f$, and the total number of events $N$.  Since we
have underlying distributions for each component, we can now make
expected distributions in the $f$-$N$ plane for each model.  We do so for
all three different hypotheses we wish to compare --
QDM + SM, GDM + SM and pure SM background.

The central values of $N$ and $f$ in the signal cases depend both on $\MDM$ and $M_*$.
We generate a number of pseudo-experiments using the above mentioned
distributions. For each pseudo-experiment we retain the value of $N$ (the total number of events), and
also obtain the best fit value for $f$. We hence get a histogram in
the $f$-$N$
plane for each of the three hypotheses.

The overlap of these two histograms is indicative of sensitivity of
the experiment to the specific value of $M_*$. Quantitatively, the
expected exclusion or discovery confidence can be calculated using for
instance the
Pearson's $\chi^2$-test,
\begin{align}
  \chi^2
  &=
  \frac{
  (\mu^{N_{exp}} - \mu^{N_{obs}})^2
  }
  {(\sigma^{N_{exp}})^2
  }
  +
  \frac{
  (\mu^{f_{exp}} - \mu^{f_{obs}})^2
  }
  {(\sigma^{f_{exp}})^2 
  }
\end{align}
where $\mu, \sigma$ denote the mean and standard deviation obtained
for observed and expected (from a given hypothesis) quantities.  For
the emulation of the counting experiment, we use the $\chi^2$ value
derived from the number of events alone. Similarly, for the separation
of hypotheses, we use the $\chi^2$ value derived from only $f$.
Technically, the parameters $f$ and $N$ are correlated, so treating
them as uncorrelated over-estimates the error. However, we will see
that the correlation in the relevant region of parameter space is
small.

For exclusion at 8 TeV, we use the
number of events observed in the ATLAS
analysis~\cite{ATLAS-CONF-2012-147} for $\mu^{N_{obs}}$, and assume
$\mu^{f_{obs}}$ to be at the SM prediction. We can then calculate the
$\chi^2$ statistic for excluding each model, QDM + SM, or GDM + SM.

For discovery of each model at 14 TeV, we assume for $\mu^{N_{obs}}$ and
$\mu^{f_{obs}}$ values predicted by the corresponding model. 
In this case, we calculate the $\chi^2$ statistic for excluding the SM
hypothesis. The value of $M_*$ which allows us to rule out the SM at
5$\sigma$ level is the discovery reach of the analysis.

In the next section we will use the parameters $f$ and $N$ in conjunction using the $\chi$-squared test statistic to enhance the sensitivity of monojet searches for both exclusion limit setting and discovery. In the event of a discovery of an excess in monojet searches, the operator is clearly not determined uniquely. We will also present our results for separating the QDM hypothesis from the GDM hypothesis in the event of such a discovery using the parameter $f$.




\section{Results}
\label{sec:results}
We will now present our results for the expected enhanced sensitivity of monojet searches when using JEPs.
Our results will address two different aspects of the
monojet search. We study the improvement in the exclusion limits placed by monojet searches when using JEPs at the 8 TeV LHC. We use the ATLAS 10.5 fb$^{-1}$ search \cite{ATLAS-CONF-2012-147} as a benchmark for this case.
We also consider the discovery reach at the 14 TeV LHC and study the
extent to which the two different models (QDM and GDM) of dark matter
can be separated in the event of a discovery.

\begin{figure}[tp] \begin{subfigure}[t]{0.45\textwidth}
  \includegraphics[width=\textwidth]{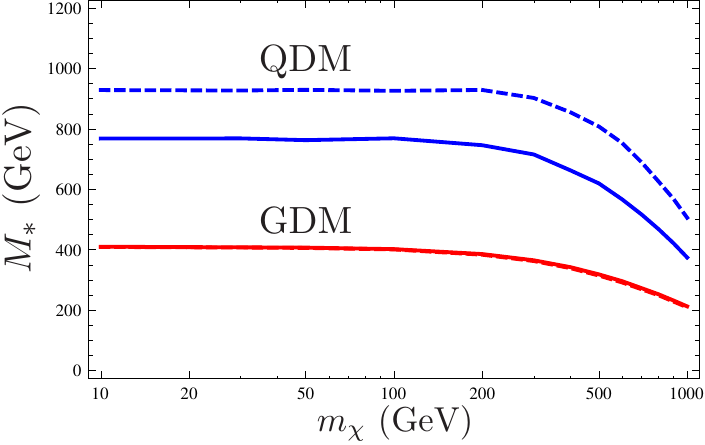}
  \caption{}
  \label{subfig:exc8}
\end{subfigure} 
\qquad
\begin{subfigure}[t]{0.45\textwidth}
  \includegraphics[width=\textwidth]{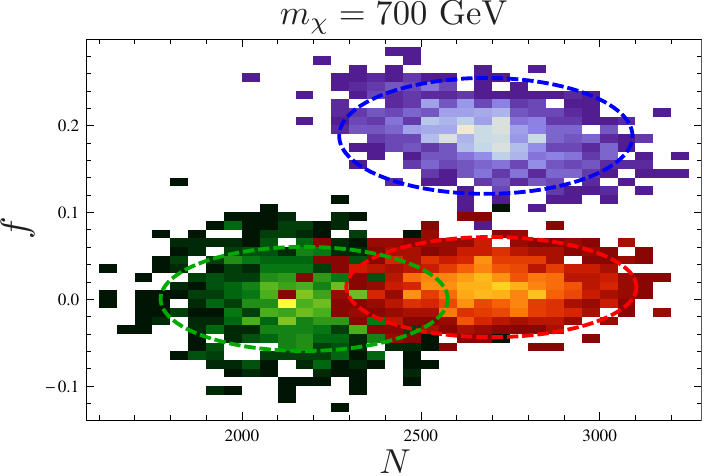}
  \caption{}
  \label{subfig:dencon8} 
\end{subfigure}
\caption{Exclusion
limits from our monojet analysis at the 8 TeV LHC at 10.5 fb$^{-1}$.
Left: $95\%$ limits on $M_*$ based on the counting experiment (solid
lines) and the joint exclusion limits using both the counts and the
JEPs (dashed lines). Limits are shown for both dark matter models,
QDM (blue) and GDM (red).  Right: A normalized histogram with
$1\sigma$ contours of pseudo-experiment distributions in the $f, N$
plane for dark matter mass $m_\chi = 700$ GeV. The hypothesis
considered are pure SM background (green), QDM + SM (blue) and GDM +
SM (red). The values of $M_*$ for each signal model were chosen to
be at the bound derived from the counting experiment (solid lines in
the left figure).} \label{fig:denexc8} \end{figure}

\subsection{Exclusions at 8 TeV LHC}

As shown in Fig.~\ref{fig:explimits}, current ATLAS and CMS analyses
put constraints on the suppression scale of various dark matter
effective operators, parameterized as $M_*$. In this section we will
show the expected improvement in the ATLAS exclusion limit when using
JEPs.

In Fig.~\ref{subfig:exc8}, we plot the 95\% exclusion limits for the
QDM and GDM models. We show the limits obtained by just the
\emph{cut-and-count} analysis of the experiment, which agrees well
with the limits set by the experimental analysis. We also present 95\%
exclusion limits obtained from a combined $\chi^2$ test using the JEPs
parameterized by $f$ and the number of events $N$. We use observed and
expected background events published in \cite{ATLAS-CONF-2012-147}. We
assume that the observed jet energy profile is consistent with the SM
prediction when setting the exclusion.

We see that for the QDM model the limits obtained are somewhat more
stringent. This is due to a strong separation in the JEP for the
background only sample and the SM + QDM sample as shown in
Fig.~\ref{subfig:dencon8}. The average JEP in the case of GDM is much
closer to the background, and hence there is no significant
improvement in the limits in this case.

The improvement in limits is moderate. This is partially due to the
fact that the current limits are set in the regime where the
statistical and the systematic uncertainty on the expected number of
background events is comparable, and $S/B \ll S/\sqrt{B}$.  Therefore,
the overall JEP of signal + background is dominated by the background
(i.e. $V+j$) profile $(f \simeq 0)$. Further, the dark matter event
rate is highly sensitive to the value of $M_*$, and therefore
including the JEP leads to only a marginal increase in the sensitivity
to this parameter.

Nonetheless, measuring the value of $f$ constitutes an independent
check of the SM background hypothesis. A value of $f$ which is
consistent with zero points to an absence of an appreciable dark
matter signal, even without a precise prediction of the background.
Therefore, in searches where the number of background events is
subject to large systematic uncertainty, the JEP measurement can
provide valuable complementary information.

\begin{figure}[tp] \begin{subfigure}[b]{\textwidth}
  \includegraphics[width=0.45\textwidth]{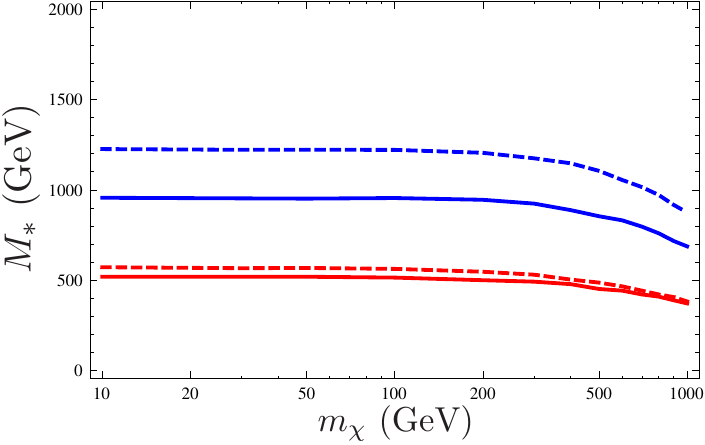}
  \qquad
  \includegraphics[width=0.45\textwidth]{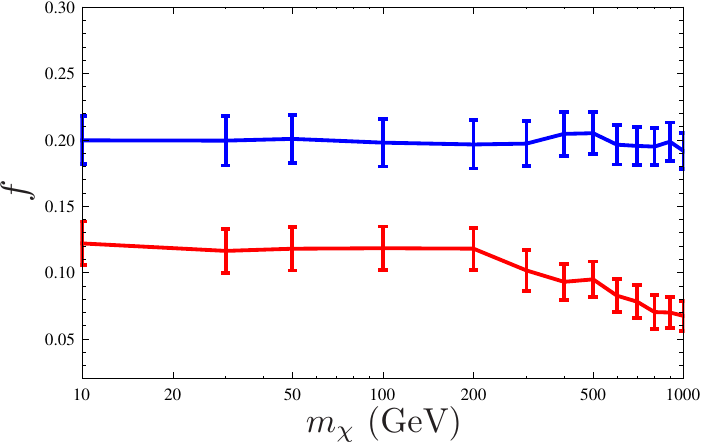}
  \vspace{2mm} \caption{10 fb$^{-1}$} \label{subfig:dissepa}
\end{subfigure} 
\\ 
\begin{subfigure}[b]{\textwidth}
  \includegraphics[width=0.45\textwidth]{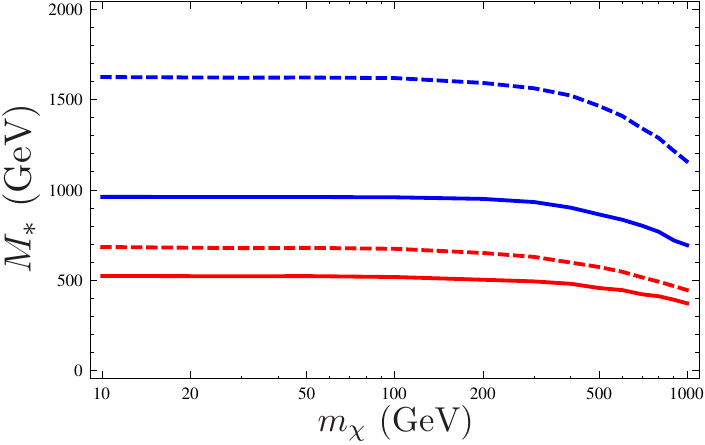}
  \qquad
  \includegraphics[width=0.45\textwidth]{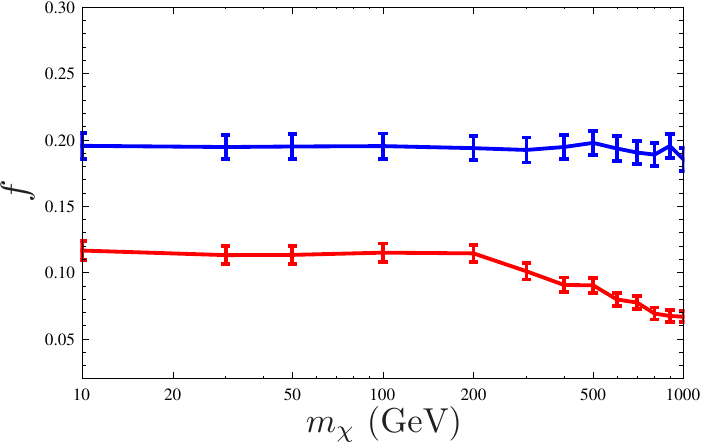}
  \caption{100 fb$^{-1}$} \label{subfig:dissepb} \end{subfigure}
  \caption{ 5$\sigma$ discovery potential and model separation at 14
  TeV at a) 10 fb$^{-1}$ and b) 100 fb$^{-1}$ with a 5\% systematic
  uncertainty on the background prediction.  Left: We show the
  discovery reach for the parameter $M_*$ based on the counting
  experiment (solid lines) and the joint discovery by both the
  counting and the jet energy profiles (dashed lines). Limits are
  shown for both operators, QDM (blue) and GDM (red). Right: We also
  show the level of separation achieved between the two operators in
  the presence of background, QDM (blue) and GDM (red), when $M_*$ is
  fixed at the 5$\sigma$ discovery limit for each mass.  }
  \label{fig:dis_sep} \end{figure}

  \begin{figure}[tp]
    \begin{subfigure}[t]{0.45\textwidth}
      \includegraphics[width=\textwidth]{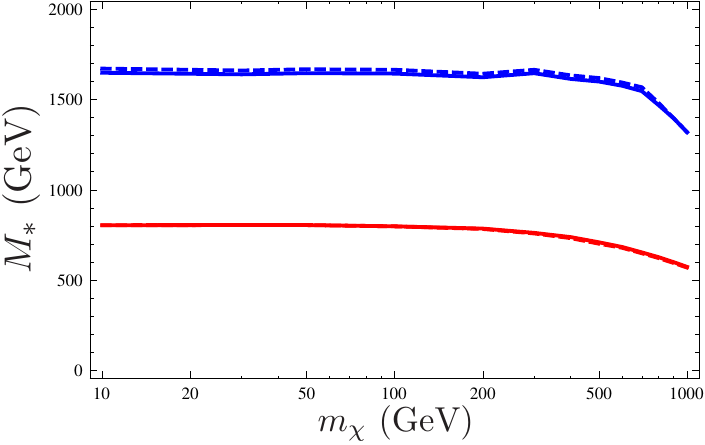}
      \caption{} 
    \end{subfigure} 
    \qquad
    \begin{subfigure}[t]{0.45\textwidth}
      \includegraphics[width=\textwidth]{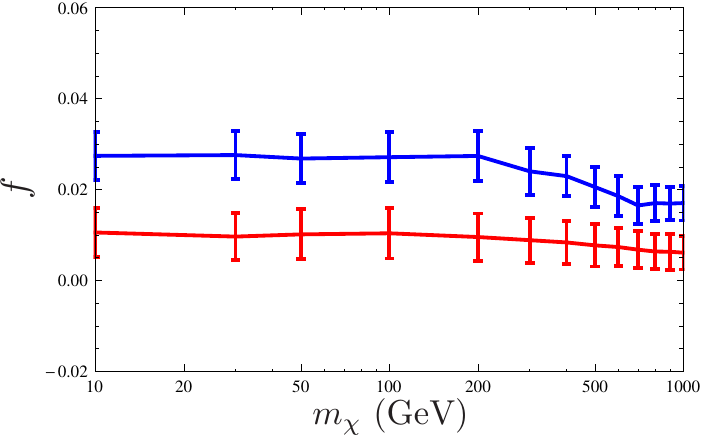}
      \caption{} 
    \end{subfigure} 
    \caption{ (a) Shown is the discovery
    reach for the operators QDM (blue) and GDM (red) in the presence of
    background at the
    14 TeV LHC at 100 fb~$^{-1}$, with only statistical errors using a
    counting experiment alone (solid lines) and with JEPs (dashed
    lines).  (b) For $M_*$ values at the discovery reach shown in (a)
    using the counting experiment alone, we show the separation in
    JEPs parameterized as $f$. We see that the QDM and GDM hypotheses
    are well separated by using the JEP.} \label{fig:dis_sep_100_0}
  \end{figure}

\subsection{Discovery reach and separation of models at 14 TeV}

We present the projected sensitivity of monojet searches at the 14 TeV
LHC at two benchmark luminosities, 10 fb$^{-1}$ and 100 fb$^{-1}$. We
use the same selection cuts as in the 8 TeV analysis.  At 14 TeV, the
production cross sections are larger by about an order of magnitude
relative to 8 TeV. Consequently, the relative statistical uncertainty
is much smaller.  We present our results with a benchmark 5\%
systematic error on the number of  background events. Note that with
the same $p_T$ cuts as in the 8 TeV analysis, the 5\% systematic error
on the number of events at $14$~TeV completely dominates the
statistical error already at 10~fb$^{-1}$ of luminosity.

We show the limits obtained by a counting experiment at the $14$~TeV
LHC in Fig.~\ref{fig:dis_sep}.  Since the systematic error dominates,
increasing the luminosity from $10$ fb$^{-1}$ to a 100 fb$^{-1}$ does
not increase the discovery reach in a counting experiment. However, in
this case the discovery reach can be greatly enhanced by including the
JEP, especially at higher luminosity. This is due to the fact that the
signal + background models show a remarkable separation in $f$ from
the background-only model even when the separation in $N$ is poor due
to systematic error dominance.  Further, the QDM + SM and the GDM + SM
hypotheses are also very well separated. This separation can be
explicitly seen in Fig.~\ref{fig:dis_sep}.

Thus, our main point is that a conventional counting experiment is
systematic dominated and would benefit immensely from the inclusion of
the jet energy profiles.  In practice, we would expect to compensate
for the large systematic error by a redefinition of the signal region
by using a larger $p_T$ cut, so as to make the statistical and the
systematic errors comparable.  In this scenario, the improvement in
discovery reach offered by the JEPs might be slightly less drastic.
For contrast, we also present results for a scenario with zero
systematic errors in Fig.~\ref{fig:dis_sep_100_0}. This is obviously
an unrealistic assumption, and we merely include it to show the
degradation of the improvement in discovery reach using JEPs. We see
that in this case the discovery reach is not improved by adding JEP.
However, even in this case we see that we can still comfortably
differentiate between the two different dark matter operators.
Therefore, the JEPs are effective and crucial in differentiating
different DM operators in monojet searches.

We reiterate here that we have not included a systematic error on the
ideal prediction of the JEP (EAP) for a given model. We expect that
our main conclusion of an improvement in the discovery reach and
separability of the DM operators will still hold true even after these
systematics are included. The detailed study of systematic
uncertainties on the JEPs is left to future work.

\section{Implications for direct detection}
\label{sec:directdet}
In this section we describe how a discovery in monojet searches would relate to predictions
for direct detection experiments.

Direct detection experiments look for recoiling nuclei from  dark matter
particles scattering off of a target. In the EFT
formalism, the same operators which give dark matter production at
the LHC (i.e.~couplings to quarks and gluons) also give rise to
scattering in direct detection experiments. Here, we shall only discuss spin-independent direct detection signals since the specific operators we have considered for QDM and GDM only give rise to these kind of interactions.

\begin{figure}[tp]
  \begin{center}
    \includegraphics[width=0.9\textwidth]{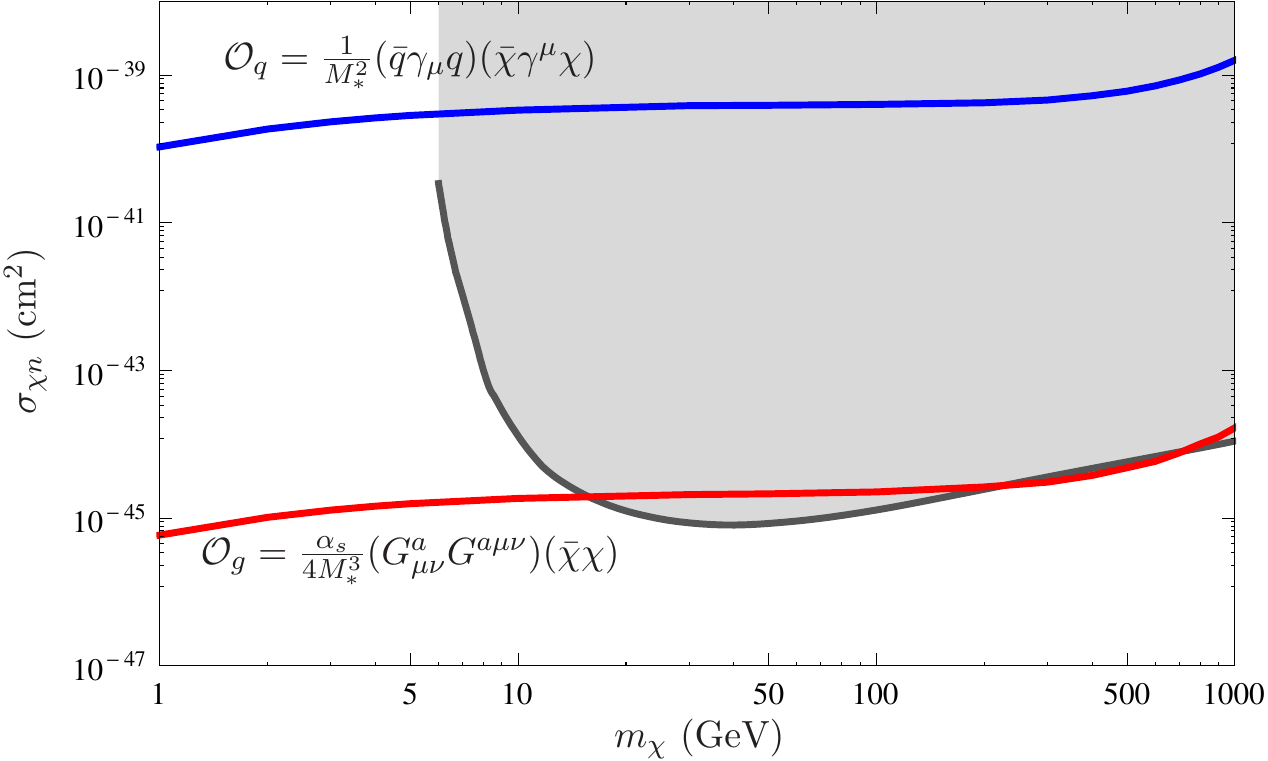}
  \end{center}
  \caption{WIMP-nucleon scattering cross section from the QDM (blue) and GDM (red)
  models described in the text. For each dark matter mass, the value
  of $M_*$ is chosen at the 5$\sigma$ discovery reach limit at the 14
  TeV LHC, by using a counting experiment with 10 fb$^{-1}$ of
  luminosity. The gray shaded region shows the recent exclusion limits from
  LUX~\cite{Akerib:2013tjd}.}
  \label{fig:ddplot}
\end{figure}

Given an EFT of dark matter scattering with quarks or gluons,
there are two parameters characterizing the EFT, $\MDM$ and $M_*$, and
signals in direct detection experiments and the LHC can be
related for each values of these parameters. However, as can be seen from Fig. \ref{fig:ddplot},
different EFT operators can give widely different predictions for
direct detection. Therefore, by the counting experiment alone, it is
not possible to directly relate signals at the LHC and direct
detection.

We have shown in this work that JEPs can be used to discriminate between the QDM and GDM EFTs at the LHC. 
This identification of whether the monojet signal is arising from a QDM or GDM EFT can have two possible implications for the connection to direct detection experiments depending upon the relevant parameter space region:

\begin{enumerate}
  \item
The dark matter model is allowed by direct detection
constraints. We can see that for the QDM signal this is only true for very
low dark matter masses, but for the GDM signal, a wide range of masses can be
consistent with direct detection bounds. Knowledge of the specific EFT responsible for the monojet signal will then
predict the rates and spectra expected at direct detection
experiments and help us to design future direct detection experiments to probe the relevant parameter space.
\item The particular model is already ruled out by direct detection,
  and hence the particle discovered in monojets at the LHC does not constitute the
  local dark matter. This then points to long-lived particles,
  multiple components of dark matter, or some other suppression of
  scattering in direct detection experiments (e.g. inelastic dark
  matter).
\end{enumerate}

Thus, the identification of the specific EFT using JEPs is a valuable and necessary tool when correlating the results of monojet searches with direct detection experiments. We note that not all EFTs can necessarily be differentiated in this way, and for a complete picture it might be necessary to use information from complementary searches.

\section{Discussion and conclusions}
\label{sec:concl}



In this paper, we studied monojet searches for dark matter at the LHC. For simplicity we focussed on 
effective field theories for dark matter interactions with
quarks (QDM) and gluons (GDM). We found that the monojet associated with each of these EFTs 
has a characteristic quark/gluon fraction initiating the jet at the parton level.
Moreover, the dominant backgrounds ($W,Z$+jets), also have a characteristic gluon
fraction, which is distinct from the signal.

We use the averaged JEP of the monojet event sample to measure the fraction of
gluon-initiated jets. While the individual event JEPs can have large
fluctuations on an event-by-event basis, the sample average JEP is
very robust, and is less sensitive to the shape of the underlying fluctuations.

The use of the averaged JEP along with the event counts provides us with better sensitivity than just using a counting experiment alone to discriminate between signal and background. We found that when the counting experiment sensitivity has large systematic errors, the additional use of JEP discrimination yields a significant improvement in our ability to set limits or discover an excess in monojets at the LHC.

In the event of a discovery of an excess in monojets, JEPs become an invaluable
tool to identify the specific dark matter interactions (separating QDM from GDM).

The EFT approach to dark matter interactions
connects signals at the LHC with direct detection experiments.
However, it does so only after the specific effective operator for
interaction between the dark matter and quarks or gluons is
identified. Different operators which give the same number of events
at the LHC can give predictions spanning five orders of magnitude at
direct detection experiments. Thus, identification of the specific QDM or GDM EFT is especially important when comparing the signals at the LHC with results from direct detection experiments.

Identification of the operator responsible for a monojet excess would sharply predict the expected signal region for direct detection experiments. A null result at direct detection experiments in the relevant region would point to the possibility that the particle discovered at the LHC is not one which makes up the local dark matter density.

We note that there are other methods that could be used to discriminate between the various DM EFTs.
Event-by-event tagging of quark-initiated and gluon-initiated jets has been proposed (see
\cite{Gallicchio:2011xq, Gallicchio:2012ez} and references therein),
and this can also be used to separate operators with different gluon fractions for the monojet.
There are also techniques beyond quark/gluon identification by
which dark matter models can be distinguished in monojet searches. Kinematic distributions of the
jet can potentially yield information about the specific dark matter operators
involved. Combining a monojet search with other searches such as mono-$W/Z$ and monophoton search can also
provide complementary information. 

For future extensions of our work it would be interesting to combine our technique with these other discrimination techniques to reach the maximum sensitivity of monojet searches.


\acknowledgments
We would like to acknowledge useful discussions with Patrick Fox, Claudia Frugiuele, Roni Harnik, Sonia El Hedri, Raoul
R\"{o}nstch, Ciaran Williams, and C.-P. Yuan.
V.R. is supported by NSF Grant No. PHY-0855561.
PA would like to acknowledge support by the
National Science Foundation under Grant No. PHYS-1066293 and the hospitality of
the Aspen Center for Physics, where a part of this work was completed.
 Fermilab is operated by Fermi Research Alliance, LLC under Contract No.
DE-AC02-07CH11359 with the United States Department of Energy.

\bibliography{monojet-tagging}
\bibliographystyle{JHEP}    

\end{document}